# Embracing advanced AI/ML to help investors achieve success: Vanguard's Reinforcement Learning for Financial Goal Planning

Shareefuddin Mohammed,  Rusty Bealer, Jason Cohen

## Executive Summary:

- In the world of advice and financial planning, there is seldom one "right" answer. While traditional algorithms have been successful in solving linear problems, its success often depends on choosing the "right" features from a dataset, which can be a challenge for nuanced financial planning scenarios. Reinforcement learning is a machine learning approach that can be employed with complex data sets where picking the right features can be nearly impossible.

- In this paper, we'll explore the use of machine learning (ML) for financial forecasting, predicting economic indicators, and creating a savings strategy. Vanguard's ML algorithm for goals-based financial planning is based on deep reinforcement learning that identifies optimal savings rates across multiple goals and sources of income to help clients achieve financial success.

- Vanguard learning algorithms are trained to identify market indicators and behaviors too complex to capture with formulas and rules, instead, it works to model the financial success trajectory of investors and their investment outcomes as a Markov decision process.

- We believe that reinforcement learning can be used to create value for advisors and end-investors, creating efficiency, more personalized plans, and data to enable customized solutions.

## Introduction:

Financial planning is a delicate blend of science and art. One side of the coin involves financial facts and figures, while the other side factors in values, goals, and discipline. As technology continues to revolutionize financial services, advances in artificial intelligence, machine learning, and computing power are enabling organizations to sort through large data sets and build learning models to support goals-based financial planning.

Financial institutions are exploring the use of AI-powered solutions in areas such as algorithmic trading, fraud detection, and crisis management. Vanguard is leveraging the power of AI to help solve business and investor challenges in support of goals-based financial planning.

# Building a financial plan

Goals-based financial planning allows clients to save for multiple financial objectives across various time horizons. When creating a financial plan, financial advisors typically consider client assets, cash flows, liabilities, asset allocation, and risk tolerance—along with economic indicators and historical fund performance—to help a client navigate their options.  Each plan is personalized and closely monitored to ensure it accurately captures client goals and the current economic environment.

These challenges can be captured by Markov Decision Processes: We have a cash-flow modeling environment which our agent is interacting with. The agent is able to observe what happens to income projections, and likelihood of achieving financial planning goals when we change savings allocations. The agent then receives rewards in response to its actions, and the agent seeks to maximize the reward received. Vanguard's machine learning model for goals-based financial planning seeks to determine the optimal savings and investment strategy that optimizes the likelihood of achieving multiple goals.

**Reinforcement learning for financial planning**

Vanguard created a machine-learning model to provide financial advisors with insights that can then be used with clients to make financial planning decisions to optimize success. The model uses Vanguard Asset Allocation Model (VAAM), our proprietary quantitative model for determining asset allocation among active, passive, and factor investment vehicles. The VAAM framework is similar to Harry Markowitz's mean variance optimization for portfolio construction—a concept that seeks to generate the highest possible return based on a chosen level of risk—but with an additional layer that recognizes the impact of behavioral finance.

Vanguard's machine learning model elevates Markowitz's work, taking into consideration the four component goal-based wealth management framework proposed and developed by (Chhabra, 2005), (Das, et al., 2010), and (Das, et al., 2018) and simultaneously optimizes across the three dimensions of risk-return trade-offs (alpha, systematic, and factor). VAAM incorporates Vanguard's forward-looking capital market return and client expectations for alpha risk and return to create portfolios consistent with the full set of investor preferences, solving for portfolio construction problems conventionally addressed in an ad hoc and generic manner. It assesses risk and return trade-offs of portfolio combinations based on user-provided inputs such as risk preferences, investment horizon, and parameters around asset classes and active strategies.

As VAAM involves executing recurring and rule-based processes involving variable inputs and then making sequential decisions with uncertainty—making it a great application for reinforcement learning. Vanguard's initial objective was to train the model to learn the value function that maximizes the expected return with one retirement goal, several pre-retirement goals, and a debt pay-off goal. The asset allocation problem was modelled as a Markov decision problem.

Building ML pipelines
A machine learning pipeline describes or models a ML process, such as writing code, releasing it to production, performing data extractions, creating training models, and tuning the algorithm. Using cloud-based technology, Vanguard developed reinforcement learning pipelines that interact with the environment, generate observations, and learn from experience. The pipelines were designed to autonomously explore and optimize thousands of possible decisions to effectively meet pre-retirement financial goals, like purchasing a car in 2030, while also ensuring success in retirement. The reinforcement learning agent was trained using a Proximal Policy Optimization (PPO) algorithm to make decisions based on the investor's current level of wealth, income bracket, spending level etc. taking into consideration all four elements of the goal-based advice (Das, et al., 2018).

## Preliminaries:

In order to formally introduce reinforcement learning, we will first describe a discrete time stochastic control problem known as a Markov decision process (MDP). An MDP is specified as a tuple $\langle S, A, P, R, \gamma \rangle$. Here $S$ represents the state space of the underlying dynamical system and $A$ the set of permissible actions, also known as the action space. The term $P$ describes the transition probability of the underlying Markov chain of states such that

$$P^a_{ss'} = P[s_{t+1} = s'|S_t = s, A_t = a]. \tag{1}$$

The term $R$ represents the reward function of the MDP. The expected reward for taking action $a$ at state $s$ given by

$$R^a_s = E[R_{t+1}|S_t = s, A_t = a]. \tag{2}$$

The term $\gamma$ is called the discount factor and represents how much a decision maker values immediate rewards in comparison to future rewards when interacting with the above system. It is a positive real number in the range [0,1]. For an infinite horizon MDP, the return $G_t$ is defined as the cumulative discounted sum of rewards,

$$G_t = \sum_{k=0}^{\infty} \gamma^k R_{t+k+1}. \tag{3}$$

Clearly, $\gamma = 1$ represents the scenario in which all future rewards are valued equally when computing the return whereas $\gamma = 0$, only the immediate reward is included. The MDP is a highly general and useful formalism to describe sequential decision making problems that involve uncertainty. In the context of MDP, the agent is defined as the entity that makes the decisions and the environment as the dynamical system whose behavior the agent aims to control. Given that the state at time $t$, $S_t = s$, the agent interacts with the environment according to a policy $\pi(a|s)$ which is defined as the conditional probability of the agent choosing action $a$ given $s$. At this stage, the environment transitions into a new state $s_{t+1}$ according to the transition density $P^a_{ss'}$ and emits a reward according to the reward function $R^a_s$. The interaction between a deep RL agent and the environment given a policy $\pi(a|s)$ is depicted in Figure 1. The transition density of the resulting Markov chain is given by $P^\pi_{ss'}$, where

$$P^\pi_{ss'} = \sum_{a \in A} P^a_{ss'} \pi(a|s) \tag{4}$$

Similarly the expected reward at $s$ from adopting the policy $\pi(a|s)$ can be calculated as

$$R^\pi_s = \sum_{a \in A} R^a_s \pi(a|s). \tag{5}$$

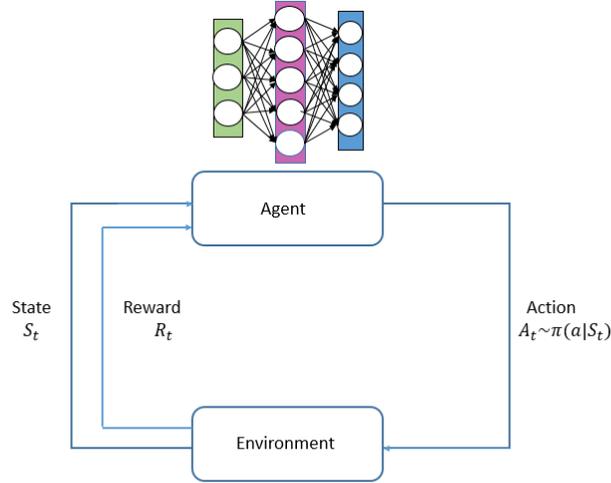

*Figure 1: Interaction between the deep RL agent and environment*

The equations (4), (5) show how both the time evolution of the system and the reward sequence are affected by the choice of policy function. The objective of the agent in an MDP is to obtain a policy $\pi^*(a|s)$ such that the expected return $E[G_t]$ is maximized. In order to formalize the optimization objective, we shall define the state value function $V_\pi(s)$ and action value function $Q_\pi(s,a)$. The state value function $V_\pi(s)$ is defined as the expected return starting from state $s$ and following policy $\pi$,

$$V_\pi(s) = E_\pi[G_t|S_t = s]. \qquad (6)$$

The action value function $Q_\pi(s,a)$ is defined as the expected return from taking the action $a$ at state $s$ and then following the policy $\pi$.

$$Q_\pi(s,a) = E_\pi[G_t|S_t = s, A_t = a]. \qquad (7)$$

Define the optimal action value function $Q^*(s,a)$ as the maximum action value function over policies

$$Q^*(s,a) = \max_\pi Q_\pi(s,a). \qquad (8)$$

Then an optimal policy $\pi^*(a|s)$ is one that achieves the optimal action value function. Given $Q^*(s,a)$, a deterministic optimal policy to the MDP can be found as

$$\pi^*(a|s) = \begin{cases} 1 \ if \ a = arg.\max_{a \in A} Q^*(s,a) \\ 0 \end{cases} \qquad (9)$$

When all the elements of the MDP including the transition density and the reward function are known, the optimization problem can be solved using what is known as dynamic programming (DP). Value iteration and policy iteration are two dynamic programming algorithms that can be used to solve the MDP. However, when the state space and action space becomes large, these computational cost of these solutions become untenable. Reinforcement learning is an alternative approach to solve the MDP that does not require the explicit knowledge of the transition probability density or other elements of the MDP. It learns from experience obtained through forward simulations of the system. Unlike DP algorithms, it does not require calculation of the value function over all states and actions at all times. Depending on whether the agent learns only the value action, the policy function or a mix of both, RL algorithms are classified into value based approaches, policy based approaches, and actor critic methods. In the present work, we have adopted a value function based RL algorithm known as DQN. The DQN agent learns the action value function that maximizes the expected return. It relies on a neural network approximation of the action value function learned through Q learning. DQN uses the temporal difference approach to update the Q-function, i.e., at state $S_t$, it picks action $A_t$ according to

$$A_t = arg.\max_{a \in A} Q(S_t, a). \tag{10}$$

Then once the system transitions into new state $S_{t+1}$ and the reward $R_t$ is observed, the Q-function is updated as

$$Q(S_t, A_t) = Q(S_t, A_t) + \alpha \left( R_t + \gamma \max_{a \in A} Q(S_{t+1}, a) - Q(S_t, A_t) \right) \tag{11}$$

Here $\alpha$ is the learning rate. The DQN algorithm uses experience replay and periodic updates to stabilize learning process. In order to obtain better coverage of the state space, we use an $\epsilon$ – greedy approach to training the DQN agent where the current action $A_t$ is chosen greedily as in eq with probability (1- $\epsilon$). The action is chosen at random with probability $\epsilon$.

# Multi Goal Financial Planning Problem:

The objective of a multi goal financial planning problem is to obtain the optimal financial strategy for an individual to meet multiple pre-retirement financial goals and be successful in retirement. Meeting a pre-retirement financial goal requires that the investor is able to assemble a specified threshold level of funds towards meeting that goal. Hence a pre-retirement goal $U_k, k \geq 1, \ k \in N$ is specified in terms of a goal target amount $H_k$ and target year $T_k$. A retirement goal $U_0$ on the other hand is specified in terms of a post in terms of target retirement year $T_o$ and the post-retirement annual spending level $H_0$. A financial strategy for meeting a pre-retirement goal $U_k$ is considered successful if the probability of meeting the goal target amount $H_k$ exceeds a specified target probability $P_k$. We shall call this the target success rate associated with goal $U_k$. For retirement goal, a strategy is considered successful if the probability of meeting the spending level $H_0$ falls within a range $[P_0, P_0 + \Delta]$. Here, an upper bound on the success probability based on the tolerance level $\Delta$ is specified to avoid overly conservative strategies that significantly penalize pre-retirement quality of life in order to achieve a large success rate for post-retirement spending level. In this work, we have utilized the same threshold probability for all goals, i.e., $P_0 = P_k$, $k \geq 1, k \in N$.

In order to obtain the optimal contribution strategies using RL, we model the multi-goal financial planning problem as an MDP. To this end, we designed the state, action, reward etc. as described below.

- **State:** State should describe all relevant information that the agent requires to predict the future behavior of the system based on what it has already observed. We describe the state in terms of a combination of static and dynamic variables. This includes demographic information such as the state of domicile of the individual within the US, financial variables such as taxable, tax-free and tax deferred balances etc. Additionally we also include the total contribution amount and goal specific variables such as the number of years left until the target year, goal target amount etc. as part of the state vector.

- **Action:** Based on the income level, pre-retirement annual spending level, personal savings etc., we determine the maximum possible annual contribution $C_{max}$ an individual can make towards their financial goals. In theory, any contribution amount in the range $[0, C_{max}]$ constitutes a valid action. However, in order to simplify the problem, we discretize the $[0, C_{max}]$ range in 5% increments of $C_{max}$. The resulting action space consists of a set of 21 possible actions.

- **Reward:** The reward signal for the multi goal planning problem is designed to ensure that maximizing the expected return will lead to the agent meeting the specified financial goals. Given the action sequence $\{a_0, a_1, \ldots a_{T_{k-1}}\}$, the actual success probability $P'_k$ of an agent meeting a pre-retirement goal target amount $H_k$ is computed using Monte Carlo simulations. For pre-retirement goals, the objective of the reward $R_k$ for the goal target year $T_k$ is calculated as

$$R_{T_k} = \begin{cases} \rho_k & if\ P'_k \geq P_k \\ \rho'_k (P'_k - P_k) & otherwise \end{cases} \quad (12)$$

Here both $\rho_k$ and $\rho'_k$ are scalar constants. For the retirement year, the reward is calculated as

$$R_{T_0} = \begin{cases} \rho_o & if\ P'_o \in [P_o, P_o + \Delta] \\ \rho'_o (P'_o - P_o) & if\ P'_o \leq P_k \\ \rho'_o (P_o + \Delta - P'_o) & if\ P'_o \geq P_o + \Delta \end{cases} \quad (13)$$

The actual success rate $P'_o$ in meeting the retirement spending target amount $H_o$ is also computed using Monte Carlo simulations. The reward $R_T = 0$ for any year $T$ where $T \notin \{T_0, T_1, \ldots\}$. As a result, multi goal planning is solved as a sparse reward MDP.

- **Discount Factor:** A discount factor of 0.95 was used in calculating the returns.

- **State Transition/ Dynamics:** The transition density for the underlying MDP is not explicitly modeled. Instead, we use a forward simulator to model the environment dynamics. In forward simulations, the static demographic variables are not updated from year to year. Financial variables are updated based on the income level, spending, savings, annual contribution etc. Components of the state vector that are affected by the stochasticity of the market are updated using a proprietary Monte Carlo simulator.

# Simulations and Results:

In order to train the reinforcement learning model, we first built a custom environment using OpenAI Gym that incorporates the state transition dynamics and reward function as described in the previous section. State is modeled as a 17 dimensional vector which consists of categorical variable describing the state of domicile and the number of custom goals. The action space consists of 21 possible discrete values. The threshold success probability for pre-retirement goals is set at 70%. For retirement goal, the tolerance level $\Delta$ is set to be 6%. We use DQN algorithm to implement our agent. We used the Ray framework to manage the interaction between the DQN agent and the multi goal planning environment. To facilitate exploration of the state and action spaces, a linear $\epsilon$ −greedy schedule in which the value of $\epsilon$ decays from 1 to 0.01 over 100,000 time steps is specified. The investor profile and goal parameters are specified as the input to the training job. Subsequently, an agent is trained to learn the optimal contribution strategy for a single customer profile using Amazon SageMaker RL for 6,000 episodes. The accumulated reward obtained by the RL agent over each training episode is presented in Figure 2 Accumulated reward during training. The accumulated reward is seen to show large random variations at the beginning of the training simulation. During this stage, the agent tends to favor exploration over exploitation as the value of $\epsilon$ is relatively high. However, as $\epsilon$ decays, it gradually switches to exploitation mode wherein actions are chosen greedily using the Q-value function it has learned. At this stage, the agent is seen to converge to a strategy that gathers a large positive reward during each episode.

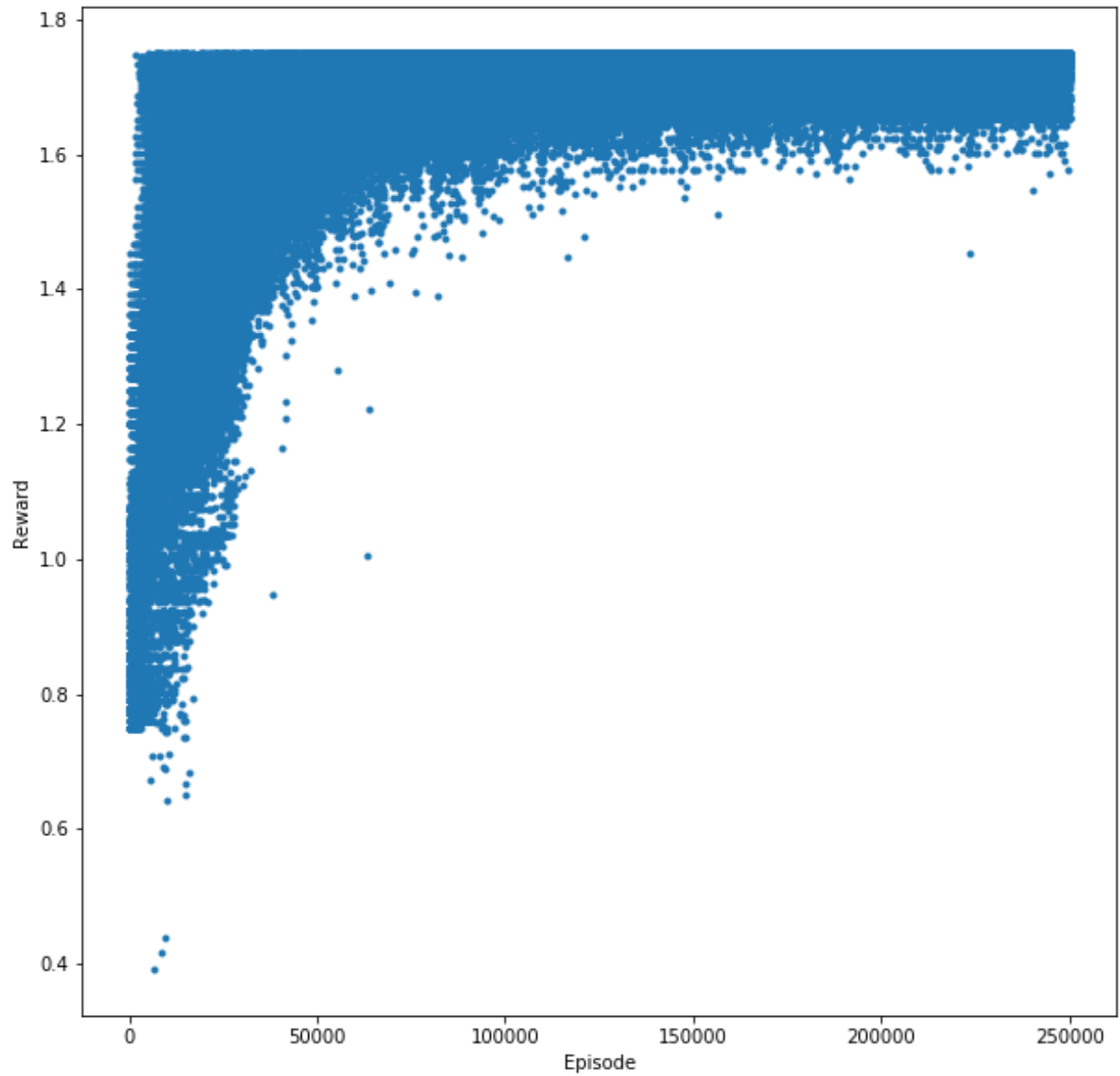

Figure 2 Accumulated reward during training

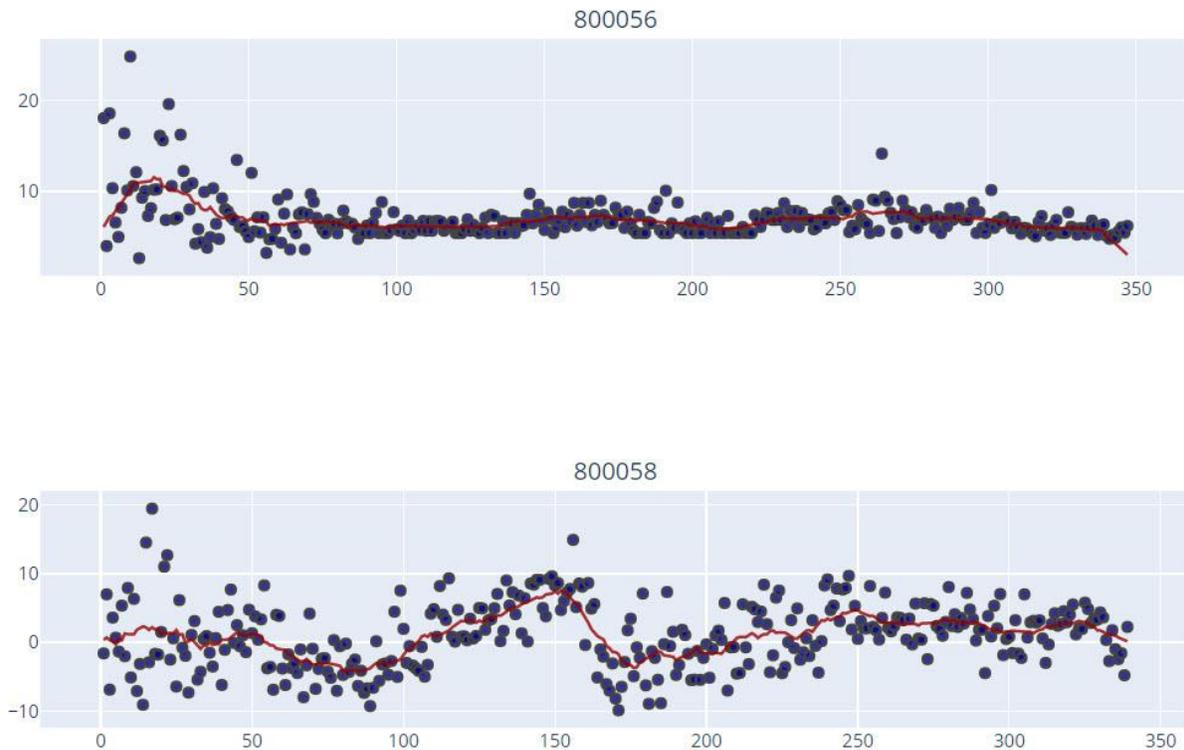

Figure 3 Moving average of Success rate convergence during training for different personas

# Conclusions:

How to determine a client's optimal asset allocation is an important sequential decision-making problem. In this paper, we demonstrated how this multi-goal financial planning problem can be modeled as a Markov decision problem. We demonstrate how a model-free reinforcement learning agent can be trained to arrive at the optimal contribution strategy to meet a retirement goal and multiple pre –retirement goals. For pre-retirement goals, the objective of the agent was to ensure that the probability of success is above a threshold (6%). While for the retirement goal, an agent is rewarded if the probability of success falls within a pre-specified range. As a result, the agent converges on a contribution strategy with a large net positive reward. Going forward, Vanguard seeks to explore the use of a reinforcement learning agent that also incorporates debt repayment as part of our multi-goal planning technique for helping investors achieve success.

Note:
All investing is subject to risk, including the possible loss of the money you invest. Be aware that fluctuations in the financial markets and other factors may cause declines in the value of your account. There is no guarantee that any particular asset allocation or mix of funds will meet your investment objectives or provide you with a given level of income.

# References

**AlmeidaTeixeira Lamartine and Oliveira Adriano Lorena Ináciode** A method for automatic stock trading combining technical analysis and nearest neighbor classification [Journal]. - [s.l.] : Expert Systems with Applications, 2010. - 10 : Vol. 37.

**Chhabra Ashvin B.** Beyond Markowitz: A Comprehensive Wealth Allocation Framework for Individual Investors [Journal]. - [s.l.] : Journal of Wealth Management, 2005. - 4 : Vol. 7.

**Cumming James** An Investigation into the Use of Reinforcement Learning Techniques within the Algorithmic Trading Domain [Report]. - [s.l.] : Masters Thesis, Imperial College London, 2015.

**Das Sanjiv R [et al.]** A New Approach To Goals Based Wealth Management [Journal]. - [s.l.] : Journal Of Investment Management, 2018. - 3 : Vol. 16.

**Das Sanjiv R [et al.]** Portfolio Optimization with Mental Accounts [Journal]. - [s.l.] : Journal of Financial and Quantitative Analysis, 2010. - 2 : Vol. 45.

**Das Sanjiv R. and Varma Subir** Dynamic Goals-Based Wealth Management [Journal]. - [s.l.] : Journal Of Investment Management, 2020. - 2 : Vol. 18.

**Huang Zan [et al.]** Credit rating analysis with support vector machines and neural networks: a market comparative study [Journal]. - [s.l.] : Decision Support Systems, 2004. - 4 : Vol. 34.

**Markowitz Harry** Portfolio Selection [Journal] // The Journal of Finance. - 1952. - pp. 77-91.

**Merton Robert C.** Lifetime Portfolio Selection under Uncertainty: The Continuous-Time Case [Journal] // The Review of Economics and Statistics. - 1969. - pp. 247-257.

**Merton Robert C.** Optimum consumption and portfolio rules in a continuous-time model [Journal] // Journal of Economic Theory. - 1971. - pp. 373-413.

**Mnih V., K. Kavukcuoglu, D. Silver, A. Graves, I. Antonoglou, D. Wierstra, and** Playing Atari with Deep Reinforcement Learning [Online]. - https://arxiv.org/abs/1312.5602.

**Nevmyvaka Yuriy, Feng Yi and Kerns Michael** Reinforcement Learning for Optimized Trade Execution [Conference] // Proceedings of the 23rd international conference on Machine Learning. - [s.l.] : ACM, 2006.

**P.Schumaker Robert and HsinchunChen** A quantitative stock prediction system based on financial news [Journal]. - [s.l.] : Information Processing & Management, 2009. - 5 : Vol. 45.

**Sam Maes Karl Tuyls , Bram Vanschoenwinkel , Bernard Manderick** Credit Card Fraud Detection Using Bayesian and Neural Networks [Conference] // Proceedings of the First International NAISO Congress on Neuro Fuzzy Technologies. - 2002.

**Silver D., J. Schrittwieser, K. Simonyan, I. Antonoglou, A. Huang, A. Guez, T. Hubert,** Mastering the game of Go without human [Journal]. - [s.l.] : Nature , 2017. - Vol. 550.

**Sutton Richard S and Barto Andrew G.** Reinforcement Learning: An Introduction [Book]. - Cambridge, MA : MIT Press, 2018.